\documentclass[preprint, superscriptaddress, floatfix, citeautoscript]{revtex4-1}

\usepackage{graphicx}
\usepackage{amsmath}
\usepackage{amssymb}
\usepackage{color}
\usepackage{dcolumn}
\usepackage{epsfig}
\usepackage{bm}
\begin{document}

\title{Synthesis of ultra-incompressible sp$^3$-hybridized carbon nitride}

\author{Elissaios Stavrou}
\affiliation{Geophysical Laboratory, Carnegie Institution of Washington, Washington, D.C. 20015, U.S.A.}

\author{Sergey Lobanov}
\affiliation{Geophysical Laboratory, Carnegie Institution of Washington, Washington, D.C. 20015, U.S.A.}
\affiliation{V.S. Sobolev Institute of Geology and Mineralogy, SB RAS, 3 Pr. Ac. Koptyga, Novosibirsk 630090, Russia.}

\author {Huafeng Dong}
\affiliation{Department of Geosciences, Center for Materials by Design, Institute for Advanced Computational Science, Stony Brook University, Stony Brook, NY 11794-2100, U.S.A.}

\author {Artem R. Oganov}
\affiliation{Department of Geosciences, Center for Materials by Design, Institute for Advanced Computational Science, Stony Brook University, Stony Brook, NY 11794-2100, U.S.A.}
\affiliation{Moscow Institute of Physics and Technology, 9 Institutskiy lane, Dolgoprudny city, Moscow Region, 141700, Russian Federation}
\affiliation{School of Materials Science, Northwestern Polytechnical University, Xi'an,710072, China}

\author{Vitali B. Prakapenka}
\affiliation{Center for Advanced Radiation Sources, University of Chicago, Chicago, IL 60637, USA}

\author{Zuzana Kon\^{o}pkov\'{a}}
\affiliation{DESY Photon Science, D-22607 Hamburg, Germany}

\author {Alexander F. Goncharov}
\affiliation{Geophysical Laboratory, Carnegie Institution of Washington, Washington, D.C. 20015, U.S.A.}
\affiliation{Key Laboratory of Materials Physics and Center for Energy Matter in Extreme Environments, Institute of Solid State Physics, Chinese Academy  of Sciences, Hefei 230031, China}
\maketitle

\textbf{Search of materials with C-N composition hold a great promise in creating materials which would rival diamond in hardness  due to the very strong and relatively low-ionic C-N bond. Early experimental and theoretical works on C-N compounds were based on structural similarity with binary A$_3$B$_4$ structural-types; however, the synthesis of C$_3$N$_4$  remains elusive. Here we explored an unbiased synthesis from the elemental materials at high pressures and temperatures. Using in situ synchrotron X-ray diffraction and Raman spectroscopy we demonstrate synthesis of highly incompressible \emph{Pnnm} CN compound with sp$^3$ hybridized carbon is synthesized above 55 GPa and 7000 K. This result is supported by first principles evolutionary search, which finds that \emph{Pnnm} CN is the most stable compound above 10.9 GPa. On pressure release below 6 GPa  the synthesized CN compound amorphizes reattaining its  1:1 stoichiometry as confirmed by Energy-Dispersive X-ray Spectroscopy. This work underscores the importance of understanding of novel high-pressure chemistry rules and it opens a new route for synthesis of superhard materials.}

Since the pioneering work of Liu and Cohen \cite{Liu1989} introducing a new class of highly incompressible material, search for C$_3$N$_4$ material harder than has diamond become the Holy Grail in the field of superhard materials. The original proposal was based on realization that carbon atoms in a hypothetical C$_3$N$_4$ compound in the $\beta$-Si$_3$N$_4$ structure are bound to 4 nitrogen atoms by very strong C-N bonds, which are shorter in comparison to the C-C bond in diamond (1.47  vs 1.53 \AA {} at ambient pressure) and the material is of low ionicity. In the subsequent work of Teter and Hemley \cite{Teter1996}, a number of structures with similar bonding properties have been proposed. These structures, with the exception of the graphite-like g-C$_3$N$_4$,  have the $sp^3$ and $sp^2$  bonding for C and N atoms, respectively with the formation of CN$_4$ corner sharing tetrahedra (see Fig.  1(c)) where C atoms are connected with 4 N atoms and N atoms with 3 C atoms.  The bulk moduli of these compounds have been calculated (using a DFT method) to be very high, exceeding that of diamond for the majority of structures, and, moreover, revealing  very high densities and the presence of  wide optical bandgaps ($>3$ eV). These early predictions were based on the structural similarity with the known A$_3$B$_4$ structural types, but for materials with atomic substitutions, e.g. $\alpha$- and $\beta$-C$_3$N$_4$ \cite{Teter1996} are analogues to $\alpha$- and $\beta$-Si$_3$N$_4$, respectively; also cubic C$_3$N$_4$ is analogous to Th$_3$P$_4$.

These first theoretical predictions have triggered extensive experimental and theoretical studies aiming at the synthesis of such unique covalent compounds. These attempts involved mainly the use of chemical precursors, such as for instance triazine based compounds, to synthesize C$_3$N$_4$ phases through various mechano-chemical techniques usually in the form of thin films and nanocrystals.  Following this route, the synthesis of $\alpha$- \cite{Chen1997}, $\beta$- \cite{Niu1993,Yu1994,Yin2003a} and mainly g-C$_3$N$_4$  \cite{Guo2003,Jurgens2003} had been reported but these results were subsequently disputed as the identification of their structures was often ambiguous due to the limited quantity  and heterogeneity of samples. Moreover, it appears that shock compression of the precursor materials results  to diamond-like carbon containing nitrogen only at the impurity level \cite{Komatsu1998,Liu2006}. The synthesis of a cubic C$_3$N$_4$ phase, diverse (less dense) from the theoretically predicted \cite{Teter1996} , has been reported \cite{Zinin2008} using high pressure and temperature conditions in the diamond anvil cell (DAC) experiments, but these results have  not been confirmed by other studies , and moreover, large volume press synthesis in similar conditions resulted in preferable diamond production \cite{Kurakevych2009}. In addition, although some new \textquotedblleft C$_3$N$_4$\textquotedblright phases  have been reported starting from g-C$_3$N$_4$, this could be the result of hydrogen present in starting compounds \cite{kojima2013}.

The assumption about the stoichiometry of the high-pressure hypothetic phase has been challenged by Cote and Cohen \cite{Cote1997}, who  realized that the original proposal based on the 3:4 composition does not gain the experimental support. Indeed, a priori, the most stable composition for the C-N materials at high pressures is unclear  because carbon can exist in both sp$^2$ and sp$^3$ hybridized states. As in the case of C$_3$N$_4$, initial theoretical predictions for the CN materials \cite{Cote1997,Hart2009,Kim2001} were based on known structural types of AB compounds like zincblende, rocksalt, body centered tetragonal (bct), etc. Later, a new cg-CN has been proposed \cite{Wang2010} as more stable than the previously predicted ones up to ca 100 GPa. A more recent theoretical study \cite{Wang2012}, based on an advanced structural search algorithm, proposed an orthorhombic \emph{Pnnm} structure to be the most stable one above ca 10 GPa (which is the stability threshold in respect to carbon and nitrogen) up to 100 GPa,   with the highest hardness. According to this study \emph{Pnnm} CN is thermodynamically stable, in comparison to carbon and nitrogen, above 10.9 GPa. This is further justified by the theoretical calculations of this study (supplemental Fig. 1, inset). It is noteworthy that the predicted \emph{Pnnm} structure is identical with $\beta$-InS  which has been proposed as the most energetically favored structure by Hart et al. based on “simple” DFT calculations of the known structural types.  In a very recent study using the same structural search method, Zhang et al. \cite{Zhang2014} proposed a different tetragonal crystal structure \emph{P42/m}, with very similar to $\beta$-InS structural characteristics, as the most stable one between ca. 10 and 22 GPa (inset of supplemental Fig. 1).

\begin{figure}[ht]
\includegraphics[width=130mm]{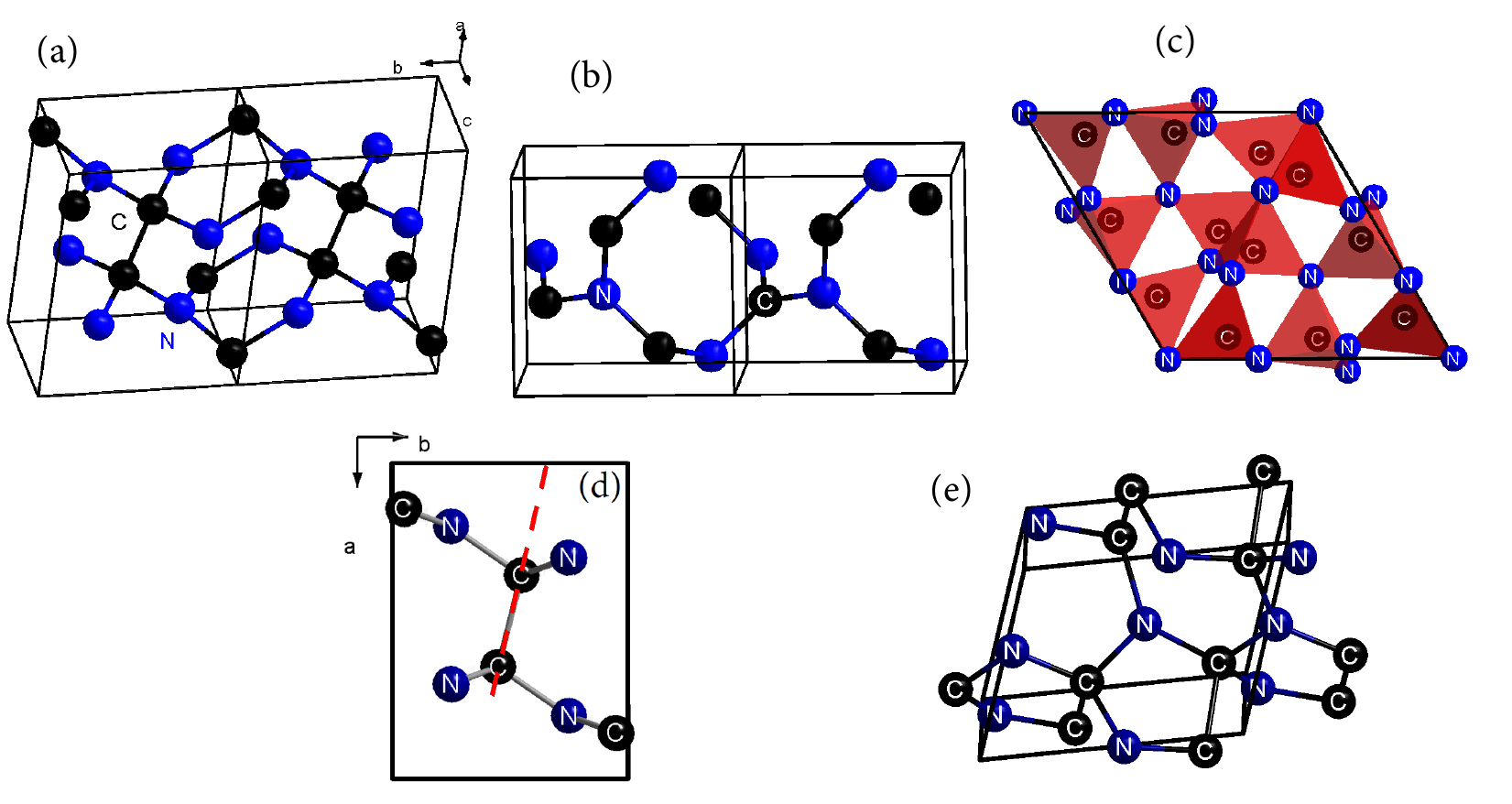}
\caption{\textbf{Schematic representations of:} (a) $\beta$-InS-type crystal structure of CN, (b) cg-CN, (c) $\alpha$-Si$_3$N$_4$-type crystal structure of C$_3$N$_4$, (d) low and high pressure $\beta$-InS-type crystal structure of CN viewed along c-axis, the tilting of the C-C bond in respect to the a axis is highlighted by the red dashed line and (e) monoclinic \emph{Pc} crystal structure of CN.  Black and blue spheres indicate carbon and nitrogen atoms respectively.}
\end{figure}

For C-N 1:1 compound composition, carbon can be either in sp$^2$ or sp$^3$ configurations. Representative phases of each family of CN materials are cg-CN and $\beta$-InS for $sp^2$ and $sp^3$ hybridized carbon respectively (Figure 1). Ultrahard materials (H $>$ 40) are expected to occur for both bonding schemes [cg-CN, $\beta$-InS], which is different from elemental carbon, which is superhard for the sp$^3$ bonding scheme only. C$_3$N$_4$ and CN compounds have a principal difference in bonding sequence: carbons are connected with only nitrogens in the former one and with three nitrogens and one carbon in the latter one. As described in details by Hart \emph{et al.}, the CN \textquotedblleft $sp^3$ family\textquotedblright arises from alternating C and N around graphite rings (see Figure 1 (a)) with additional C-C bonding between the sheets. These interlayer bonds may be between pairs of sheets, resulting in the layered GaSe-like CN structure \cite{Hart2009}, or between multiple sheets forming a 3D network as in $\beta$-InS. It is a carbon bonding configuration, which determines the material electronic bandstructure \cite{Hart2009}, that can be used as an additional sample diagnostics as will be discussed below.

According to previous theoretical studies \cite{Cote1997, Hart2009, Wang2010, Wang2012} C-N compounds with 1:1 stoichiometry are expected to be more stable, mainly due to the increased N-N separation as such bonds are not energetically favorable \cite{Cote1997}.  Additionally, the presence of strong C-C bonds  is in favor to formation   of thermodynamically  stable material. This is further justified by the theoretical results of this study (supplemental Fig. 2). However, the C-N bond length is slightly larger compared to the 3:4 compounds, thus making it necessary to thoroughly evaluate the stability and possible superhardness both experimentally and theoretically. The detailed report of the theoretical investigation of the C-N compound stability using evolutional search (USPEX code \cite{Oganov2010}) will be published elsewhere [Huafeng Dong \emph{et al.},arXiv:1408.2901v1]; here we present the results which are relevant for the major topic of the paper.

Experimentally, in this study we focus in an “unbiased synthesis” by not restricting the initial compositions of the reagents (as opposed to using chemical precursors) and also using very clean hydrogen free techniques; we use only thermodynamic stimuli (pressure and temperature) attempting to grow the most stable material irrespectively of the composition. We report the experimental synthesis of a \emph{Pnnm} ($\beta$-InS) CN phase under high pressure and high temperature conditions examined by visual observations, synchrotron X-ray diffraction (XRD), Raman confocal spectroscopy and energy-dispersive X-ray spectroscopy (EDX) probes at ambient conditions. We stress that pressure is the necessary thermodynamic stimulus for synthesis of the C-N compounds (regardless of their structure and composition) as these materials become more stable with respect to decomposition to  graphite and nitrogen above 10 GPa, mainly because molecular nitrogen is very compressible. Moreover, high temperature is essential for overcoming kinetic barriers.

We have performed numerous laser heating (LH) experiments  at various pressures  using LH systems combined with Raman spectroscopy \cite{Goncharov2009b} and XRD \cite{Prakapenka2008}. Single-crystal graphite and N$_2$ (served also as a pressure medium) have been used as reagents (see Methods). From these runs (see supplemental Fig. 1 and Table I,  for the full list of experimental runs) we observe that for pressures below 45-50 GPa and above 70 GPa diamond was the only product after quenching to room temperature. The maximum achieved temperatures were $<$ 3000 and 5000 K, respectively, as coupling of the laser radiation with the sample becomes progressively smaller once diamond forms preempting the formation of C-N compounds.  In contrast, laser heating between 50-70 GPa resulted to higher achievable temperatures ($>$ 7000 K) because of the runaway temperature increase, which we interpret as being due to the exothermic chemical reaction of C and N that occurs once the required P-T conditions are reached. No diamond formation has been observed in runs with successful C-N synthesis.

XRD patterns before LH at 55 and 65 GPa of two independent  runs   reveal 3 families of Bragg peaks: i) $\epsilon$ or/and $\zeta$ - phases of molecular nitrogen \cite{Olijnyk1990, Gregoryanz2007}, ii) from the iridium ring \cite{Cerenius2000}  used as a spacer to detach carbon sample from the diamond anvil and iii) high pressure form of carbon (supplemental Fig. 3(a), data of this study and Ref. \cite{Wangg2012}). Our LH experiments above 7000 K must be melting carbon as they reached temperatures well above the melting line of diamond at this pressure \cite{Bundy1996}, see supplemental Fig. 1. In fact, it turns out that melting carbon is a crucial precondition to synthesize CN compound.  XRD pattern of quenched samples in the close vicinity of the laser heated spot reveals the appearance of new intense Bragg peaks (Fig. 2), the decrease of the relative intensity of nitrogen related peaks and the disappearance of the broad intense HP carbon. Nitrogen peaks both before and after LH can be  indexed with the known \cite{Olijnyk1990,Gregoryanz2007} molecular phases and the determined in this study  EOS of N$_2$  is in agreement with the previous studies (supplemental Fig. 3(b)). A slight ($\approx$ 2-5 GPa) decrease of pressure was normally observed after LH. Remarkably, visual observation of the DAC indicates the presence of a transparent material exactly at the laser heated spot (inset to Figure 2). Since no Bragg and Raman \cite{Hanfland1985} peaks representative of diamond  were observed, the observation of a transparent substance   indicates the formation of a material other  than diamond.

The Bragg peaks of the new phase can be  well indexed with an orthorhombic \emph{Pnnm} (58) cell with Z=4 and a=4.77 \AA, b=3.67 \AA {} and c=2.45 \AA {} at 55 GPa (supplemental Fig. 4(a)). Moreover,  we compared the experimental and the calculated patterns of  the theoretically predicted phases with: (a) 1:1 stoichiometry, including zincblende \cite{Cote1997}, rocksalt \cite{Cote1997}, body-centered-tetragonal \cite{Cote1997}, cg-CN \cite{Wang2010}, GaSe \cite{Hart2009} and $\beta$-InS \cite{Hart2009,Wang2012} and (b) 3:4 stoichiometry, including $\alpha$-, $\beta$-, g-, cubic and pseudocubic- C$_3$N$_4$ \cite{Liu1989, Teter1996}. The almost perfect match, for all new peaks, can be observed in the case of \emph{Pnnm} $\beta$-InS-type structure  (Fig. 2 and supplemental Table II) while  there is no match with other predicted structures.   Representative Le Bail refinements of the experimentally observed diffraction patterns, after LH at the highest pressure for each run, based on the $\beta$-InS-type structure  are shown in  Figure 2 and supplemental Fig. 4 for the separate synthesis at 65 and 55 GPa, respectively. Preferred orientation  effects and strongly anisotropic peak broadening effects (Fig. 2 inset and supplemental Fig. 4(b)), which are usual in HP-HT synthesis, prevent us from a full structural refinement (Rietveld) of the positional parameters. According to the theoretical predictions both carbon and nitrogen occupy 4g (x,y,0) WP with (0.355,0.566,0) and (0.816, 0.744, 0) respectively. Using these positional parameters we observe a fair agreement between calculated and observed intensities.

\begin{figure}[ht]
\includegraphics[width=130mm]{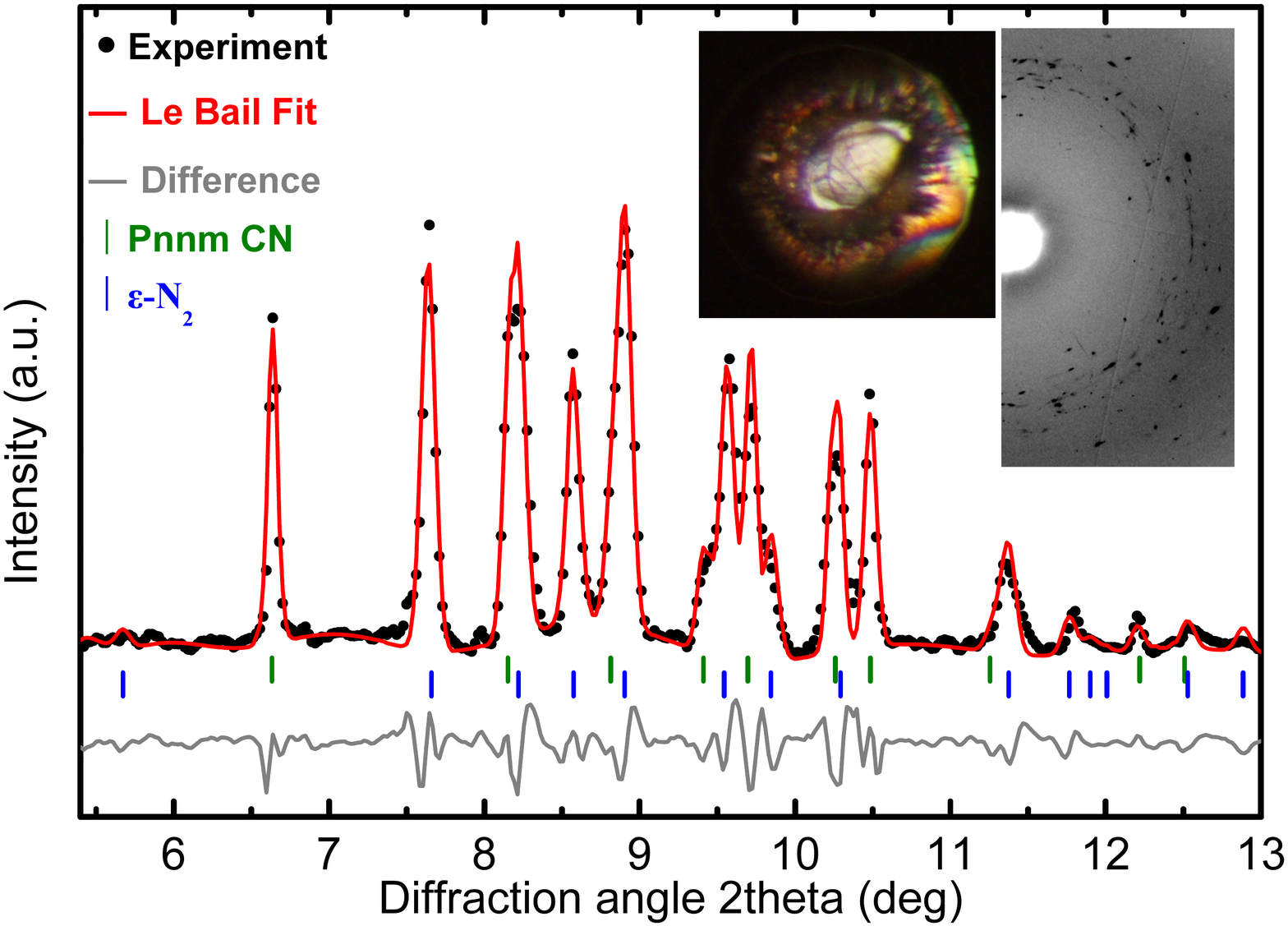}
\caption{\textbf{Le Bail refinement for CN at 65 GPa.} \emph{Pnnm} CN and $\zeta$-N$_2$ peaks are marked with
black and red vertical lines respectively. The X-ray wavelength is 0.3344 \AA. The insets show: (a)  a microphotograph of the sample after heating in transmitted light, indicating the transparency of the synthesized phase and (b) 2-D XRD image of the \emph{Pnnm} CN and $\zeta$-N$_2$ mixture.}
\end{figure}

Raman spectra (Fig. 3(a)) of the new phase show a rich spectrum, characteristic of a low symmetry cell. The lattice modes of N$_2$ \cite{Gregoryanz2007} can be readily separated as they are broader (Fig. 3).  Twelve Raman-active zone-center modes are predicted from group theory in the case of the \emph{Pnnm} structure, with the symmetries: 4$A_{g}$ + 4$B_{1g}$ + 2$B_{2g}$ + 2$B_{3g}$.  High frequency stretching modes of the CN phase are expected to be below the C-C stretching region of diamond at a given pressure, since the C-C bond in \emph{Pnnm} CN is slightly longer (1.57 \AA) than in diamond. All the observed Raman modes but one shows the increase with pressure. Above 35 GPa the high-frequency CN modes interfere with the Raman signal of stressed diamond anvils; moreover, strong sample fluorescence did not allow to obtain high quality data at these conditions. The comparison of the measured and predicted Raman frequencies versus pressure (Fig. 3(b)) shows a good overall agreement in terms of the mode frequencies, given the fact that LDA usually overestimates, while GGA \cite{Wang2012} underestimate the phonon frequencies. This documents the interatomic bonding and together with X-ray diffraction unequivocally identifies the material as \emph{Pnnm} CN.

\begin{figure}[ht]
\includegraphics[width=120mm]{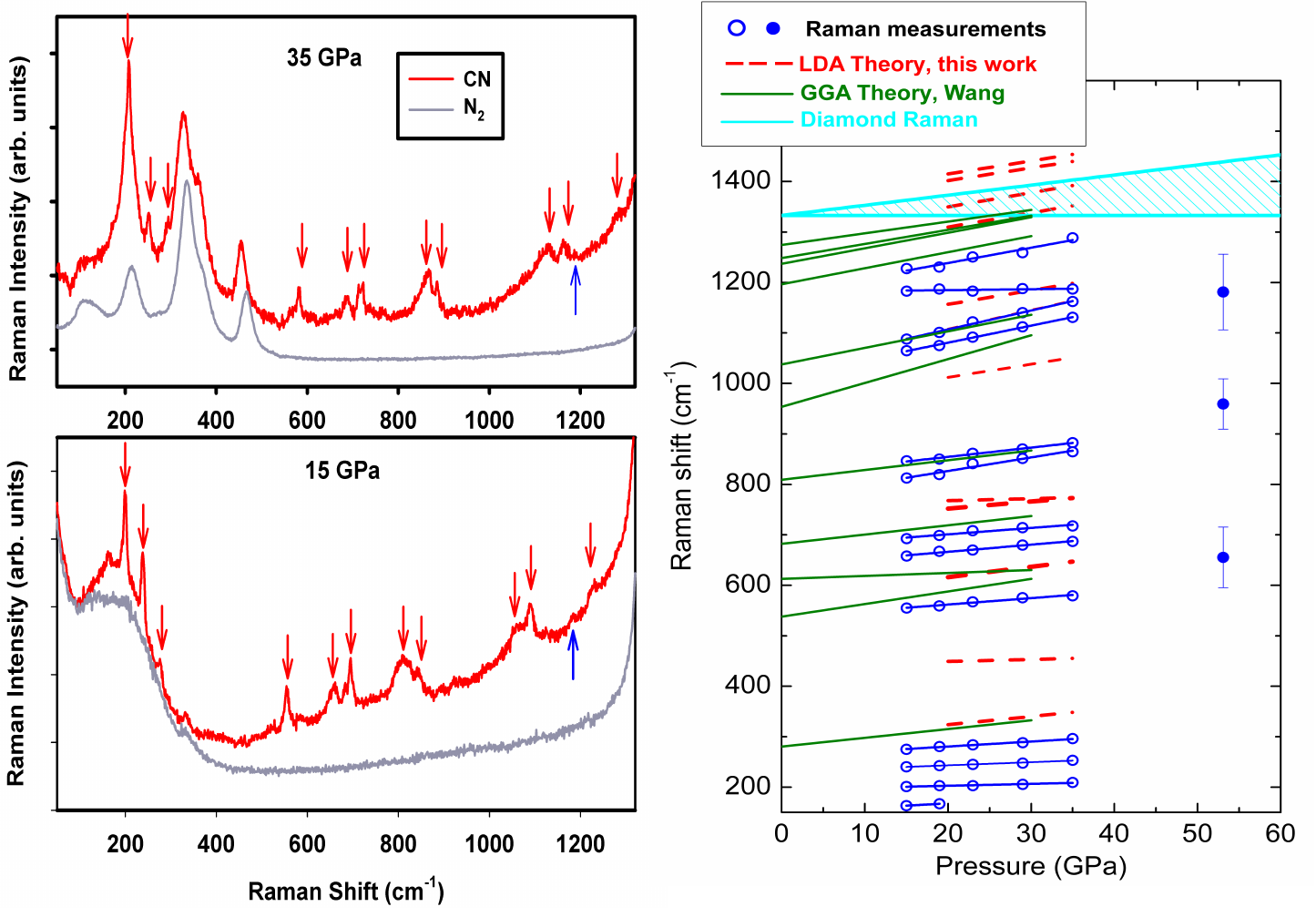}
\caption{\textbf{(a) Raman spectra  at 15 and 35 GPa.} Red curves correspond to CN and gray curves - to molecular N$_2$, measured in the same high pressure cavity. The Raman peaks of \emph{Pnnm} CN are marked by arrows: red arrows correspond to the peaks increasing (in frequency) with pressure and blue arrow pointing up- to a weak peak decreasing with pressure). (b) F\textbf{requencies of Raman modes of \emph{Pnnm} CN against pressure upon decompression.} Experimental results with blue symbols and theoretically predicted with green (GGA \cite{Wang2012}) and red (LDA this study) lines. The light blue area represents the frequency range dominated by Raman scattering of diamond. The error bars correspond to the bandwidth of broad Raman peaks, which were observed after the laser heating at 55 GPa.  Please note that only the strongest computed Raman modes are shown, which accounts for the the difference in the observed and predicted number of Raman bands.}
\end{figure}

The transparency of material in the laser heated spot, indicating a wide band gap material, is also in favor to the synthesis of the $\beta$-InS-type CN phase. Our calculations, as well as earlier theoretical predictions \cite{Wang2012}, indicate that \emph{Pnnm} CN has a wide band gap of 3.7 eV. All other predicted C-N compounds with sp$^3$ bonded carbon are also expected to be insulators \cite{Hart2009}. In contrast, cg-CN and other C-N compounds with sp$^2$ bonded carbon are expected to be metallic \cite{Wang2010}. This is in a simple analogy with graphite and diamond. Since C-C bonds should occur for sp$^3$ bonded carbon only \cite{Cote1997, Hart2009}, we conclude that the synthesized compound must have such bonds, which is an important ingredient for making this material superhard and additionally stabilizes the structure.

With decreasing pressure the newly synthesized phase remains stable up to 12 GPa as evidenced from XRD measurements and optical observations (supplemental Fig. 5).   Pressure dependence of the lattice parameters and volume per atom  equation of state (EOS) of \emph{Pnnm} CN phase together with the theoretically predicted ones are shown in Figure 4. It is interesting to note a small negative compressibility along b-axis, which  may be related to a similar observation in $\beta$-InS compound under pressure \cite{Schwarzc1995}. This observation corroborates with the Raman mode softening (Fig. 3), which is also similar to observations in Ref. \cite{Schwarzc1995}. A plausible scenario of such anisotropic behavior is that the high compressibility of a-axis results in the tilting of the C-C dumbbells,  with respect to a-axis (Figure 1(d)) so, the C-C bond distance remains almost constant.   Consequently, this results in an increase of b-axis.  We have fitted the pressure-volume data by the third-order Birch equation of state \cite{Birch1978} and determined the bulk modulus $B_0$ and its first pressure derivative $B_0'$.  The results of the fit are: $B_0$=400 (20) GPa, $B'$= 3.4 (2) and V$_0$=6.04 (3) \AA$^3$. The experimentally determined volume at ambient pressure  is in a very good agreement with theoretically predicted ones (5.98 by Hart et al. \cite {Hart2009} and 6.12 \AA$^3$ {} by Wang \cite{Wang2012}). The synthesized material appears less compressible than predicted by Hart \emph{et al.}  (no $B'$ value is given in this study so a value of 4 is used for the EOS of figure 4) but almost the same as predicted by Wang \cite{Wang2012}. So, the compressibility of the synthesized compound is comparable or even smaller than that of superhard c-BN \cite{Goncharov2007}.

\begin{figure}[ht]
\includegraphics[width=120mm]{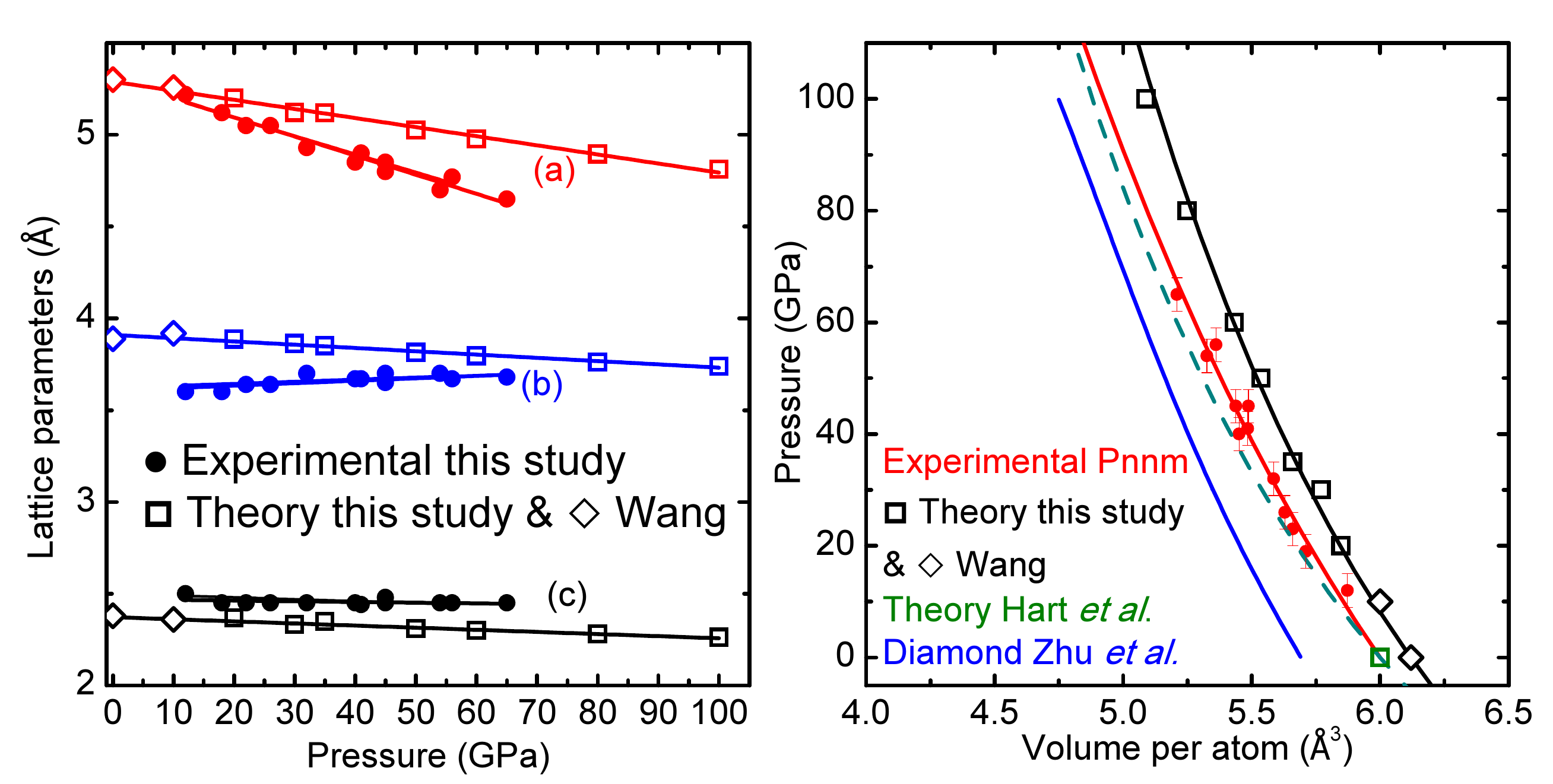}
\caption{ \textbf{Pressure dependence of (a)lattice parameters and (b)Volume per atom of \emph{Pnnm} CN.}  Experimental data are shown with solid symbols and theoretical predictions with open ones. The solid curves in (b) are the third-order Birch EOS for diamond (blue) according to Zhu \emph{et al.} \cite{Zhu2011} , the synthesized CN phase (red) and theoretical EOS predicted for \emph{Pnnm} CN  (black) according to this study and Wang \cite{Wang2012} and  according to Hart \emph{et al.} \cite{Hart2009} (green).   }
\end{figure}

It is well known that a  correlation between bulk modulus, or even shear modulus (G) which is considered more straightforward \cite{Brazhkin2004}, and hardness (H) is not direct and monotonic \cite{Gao2003}. However, a high bulk modulus is usually, but not always, a strong indication  of possible high hardness.  During the last decade, substantial effort has been made \cite{Gao2003,Keyan2008,Lyakhov2011} towards understanding the microscopic factors which determine hardness. The common consensus \cite{Oganov2014} is that high hardness depends on: (a) shortness of bond length, (b) high number of bonds, (c) high number of valence electrons and (d) high covalency of bonds. C-N 1:1 compounds satisfy all conditions and are superior in comparison to B-N due to lower ionicity \cite{Oganov2014, Keyan2008}. Moreover, since metallic bonding  always reduces hardness, due to the delocalized bonding \cite{Oganov2014}, the sp$^3$ C-N compounds (unlike metallic sp$^2$ bonded ones)  we report in this study, are the most promising ones. Indeed, the  theoretically calculated  hardness of $\beta$-InS-type CN determined as 59.6 GPa here and 62.3 GPa in Ref. \cite{Wang2012} is comparable to that of superhard cBN.  However, further studies are needed, mainly micro-hardness tests, which are difficult because the synthesized material is not recoverable at ambient pressure.

 New Bragg peaks appear  upon decompression  below 12 GPa in addition to the peaks from \emph{Pnnm} CN phase and the delta phase of N$_2$ (supplemental  Fig. 6).  These peaks cannot be attributed to the tetragonal \emph{P42/m} CN phase or the other previously proposed CN or C$_3$N$_4$ phases. We have indexed these peaks with a monoclinic cell with a volume suggestive of a 4 formula unit cell  in the case of 1:1 stoichiometry. Theoretical calculations revealed a monoclinic cell (SG \emph{Pc} (7) Z=4) which has the same bonding configuration (sp$^3$ hybridized carbon, see Figure 1) with previously proposed  \emph{Pnnm }and \emph{P42/m}  phases, albeit with higher enthalpy. With further pressure release both XRD Bragg  and Raman peaks from the synthesized compound disappear below 6 GPa and the laser heated spot becomes opaque.  EDX spectroscopy of extracted samples reveal that a CN compound  is preserved (Figure 5). We find that the stoichiometry of recovered samples, measured as a mean from several different areas, is very close to 1:1 (C$_{1}$N$_{0.82}$).  The slightly higher concentration of C could be explained due to residual (un-reacted) carbon. A plausible scenario is that  \emph{Pnnm} CN becomes unstable below ca 12 GPa with respect to decomposition to carbon and nitrogen, as predicted by the theoretical calculations of this study and also from  previous theoretical studies \cite{Wang2012, Zhang2014}. However, instead of decomposing, because of kinetic barriers,  it transforms to a metastable phase (monoclinic \emph{Pc}), while carbon remains in sp$^3$ bonding configuration. At lower pressures ($<$6 GPa) this metastable phase becomes dynamically unstable and eventually  amorphizes \cite{Deb2001,Ferlat2012} keeping its (metastable) composition.

\begin{figure}[ht]
\includegraphics[width=150mm]{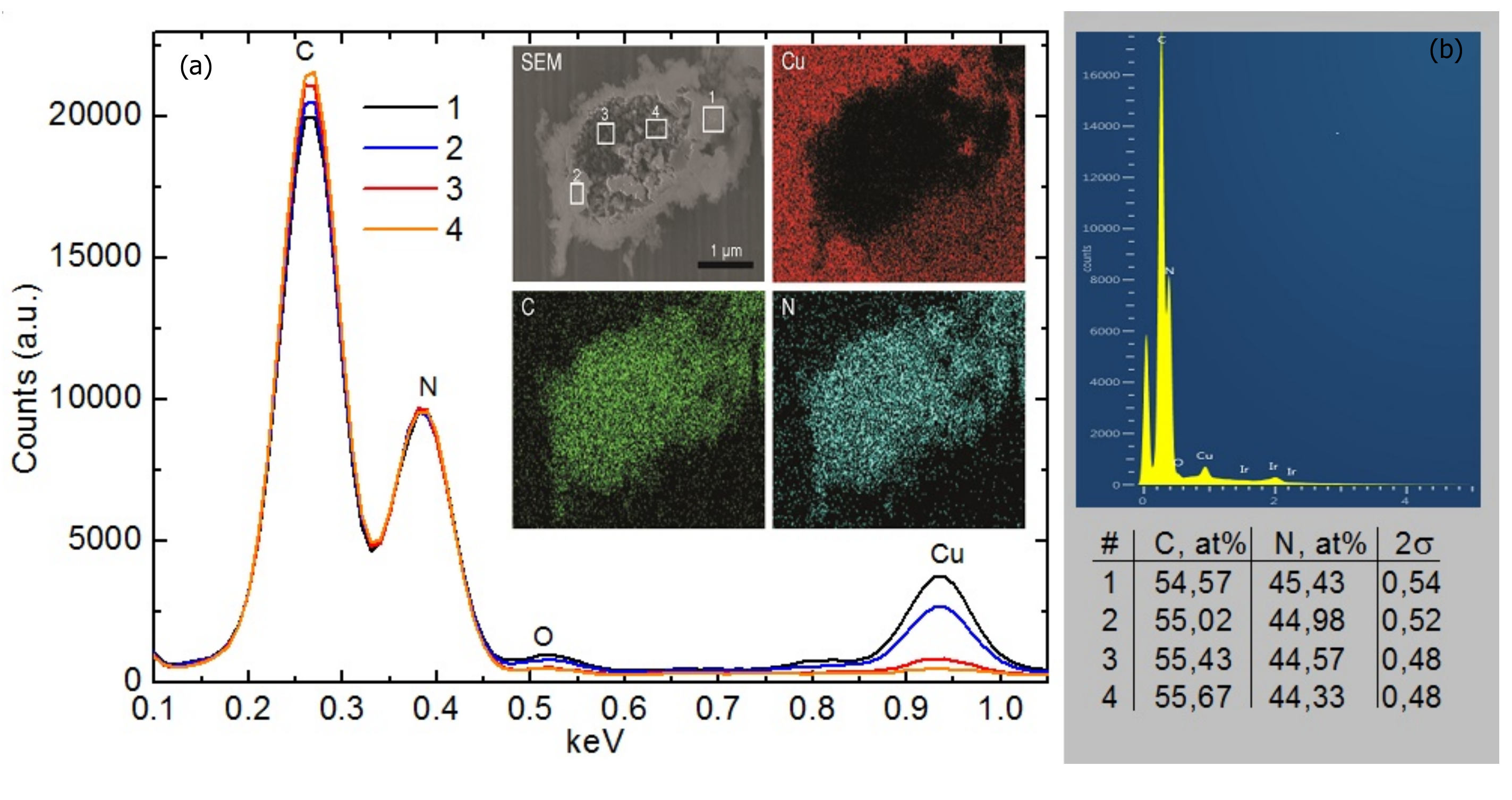}
\caption{ \textbf{(a) Representative EDX spectra, from different spots, of the CN compound at ambient conditions normalized to the N peak.} The Cu and O peaks seen from spots close to the edge arise from the Cu foil used as a conducting substrate  (see methods). The inset shows SEM/EDX micrograph and elemental maps of  the surface of the recovered CN compound. (b) representative raw EDX spectrum. The elemental analysis from different probe spots, after appropriate corrections and calibration, are summarized in the Table.}
\end{figure}

Our combined experimental study provides the first unambiguous evidence of a novel carbon nitride phase with a 1:1 stoichiometry where carbon atoms are in sp$^3$ bonding configuration forming a complete three-dimensional network. This \emph{Pnnm} CN phase was synthesized at high pressure and high temperatures in the conditions where compounds with any (presumably the most stable) composition could be formed. Other previously predicted materials (e.g., C$_3$N$_4$) did not form suggesting that they are metastable at these conditions (supplemental Fig. 2). This finding provides critical information for search of new C-N “superhard” materials with technological applications and also provides a crucial insight on the diverse chemical bonding of carbon under extreme conditions.

\bibliography{CN}

\textbf{Acknowledgments}

E.S. would like to thank Karl Syassen for providing high quality natural graphite.  This work was supported by the DARPA (Grants No. W31P4Q1310005 and No. W31P4Q1210008) and the Deep Carbon Observatory DCO. GSECARS is supported by the U.S. NSF (EAR-0622171, DMR-1231586) and DOE Geosciences (DE-FG02-94ER14466). Use of the APS was supported by the DOE-BES under Contract No. DE-AC02-06CH11357. The research leading to these results has received funding from the European Community’s Seventh Framework Programme (FP7/2007-2013) under grant agreement n 312284.
\endacknowledgments

E.S., A. G. and A. R. O. designed the study, E. S., S. L. and A. G. performed the experiments and analyzed the experimental data, H. D. and A.R.O. performed the calculations, Z. K. and V. P. performed experiments and contributed to the experimental methods.

\textbf{Methods}

Thin layers (10 $\mu$m thick) of precompressed single-crystal natural graphite fine powder were loaded in a diamond anvil cell by positioning on a top of a thin (10 $\mu$m thick) iridium or Pt ring attached to one of the diamond anvils or using a recessed gasket configuration \cite{Kolesnikov2009} (supplemental Fig. 7).    In this configuration, which ensures that graphite is detached from diamond culets, a very good thermal isolation of carbon samples was achieved. The diamond anvil cell cavity was filed with nitrogen in a gas loading apparatus, where a N$_2$ pressure about 200 MPa was created, the DAC was sealed and the pressure increased to the target pressure as determined by first order Raman mode of diamond, $\nu_1$ N$_2$ Raman vibron frequency, and Ir(Pt) EOS \cite{Cerenius2000}.

Raman studies were performed using 488, 532 and 660 nm lines of a solid-state laser in the backscattering geometry. The laser probing spot dimension was 4 $\mu$m. Raman spectra were analyzed with a spectral resolution of 4 $cm^{-1}$ using a single-stage grating spectrograph equipped with a CCD array detector. Ultra-low frequency solid-state notch filters allowed to measure the Raman spectra down to 10 cm$^{-1}$.  Laser heating was performed in a double-sided laser heating system combined with a confocal Raman probe \cite{Goncharov2009b} and also at the Extreme Conditions XRD Beamline P02.2 at DESY (Germany) and at GeoSoilEnviroCARS, APS, Chicago, which have online laser heating capabilities. Temperature was determined spectroradiometrically.  The X-ray probing beam size was about  2-4 $\mu$m.

Chemical composition of the synthesized samples was analyzed with a dual beam focused ion beam / scanning electron microscope (FIB/SEM Zeiss Auriga 40) equipped with an Oxford X-Max 80 mm$^2$ large-area silicon drift detector at 5kV accelerating voltage in the Geophysical Laboratory Carnegie Institution of Washington. Five samples of a C-N material extracted out of the diamond anvil cell were put on a Cu foil glued to an Al stub to avoid any carbon contamination. All samples were analyzed in 5-10 different spots to check for homogeneity and to account for the non-flat sample geometry, ZAF corrections have been applied to raw spectra. Ir sputter coating (ca 1nm) was used to prevent specimen charging. Boron nitride and graphite were used as standards to analyze-calibrate nitrogen and carbon content, respectively.

Predictions of stable phases were done using the USPEX code in the variable-composition mode \cite{Oganov2010}. The first generation of structures was produced randomly and the succeeding generations were obtained by applying heredity, atom transmutation, and lattice mutation operations, with
probabilities of 60$\%$, 10$\%$ and 30$\%$, respectively. 80$\%$ non-identical structures of each generation with the lowest enthalpies were used to produce the next generation. All structures were relaxed using density functional theory (DFT) calculations within the Perdew-Burke-
Ernzerhof (PBE) \cite{Perdew1996}, as implemented in the VASP code \cite{Kresse1996}.

\clearpage

\textbf{Supplemental Material}

\setcounter{figure}{0}

 \begin{figure}[ht]
\includegraphics[width=130mm]{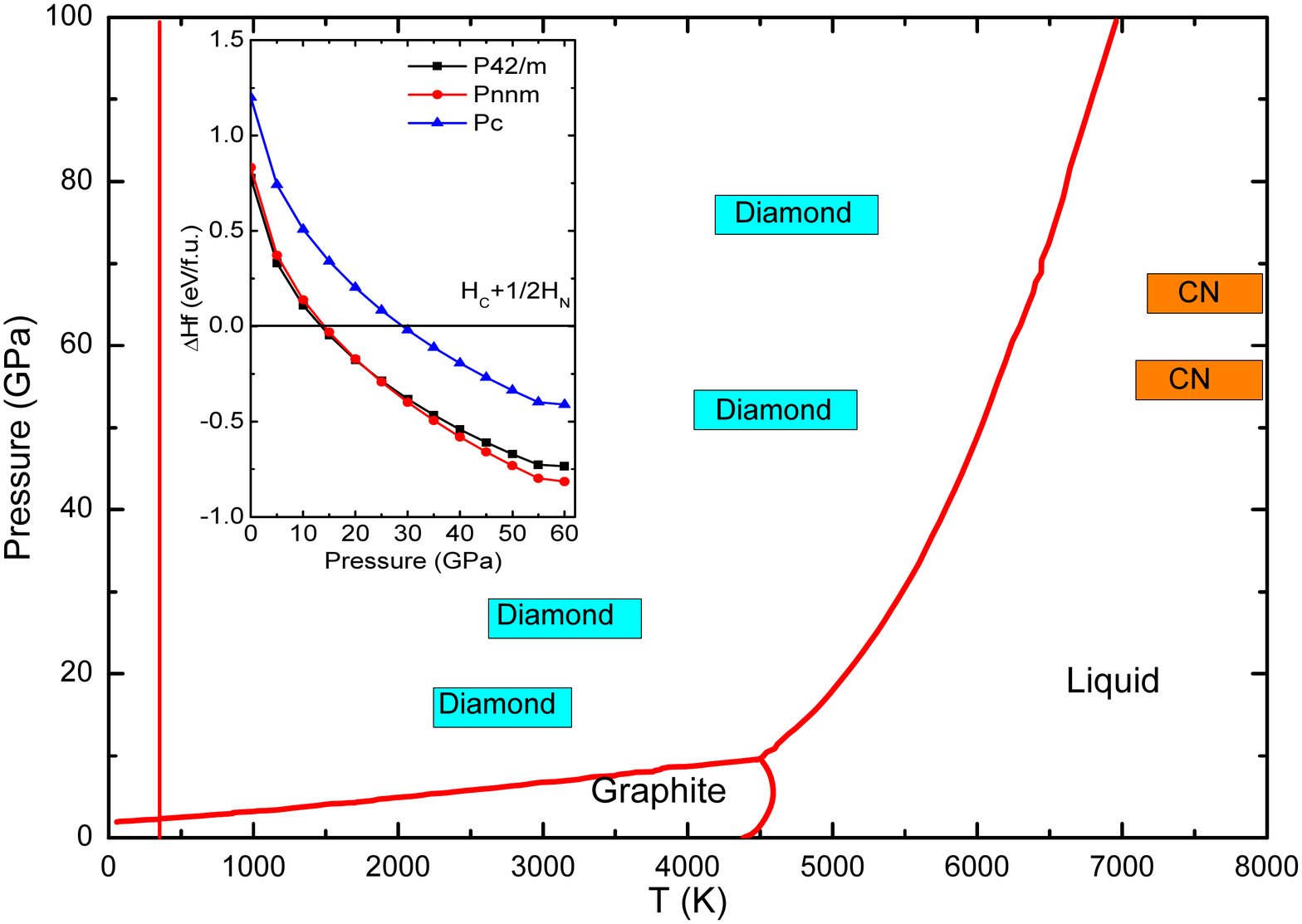}
\caption{\textbf{Pressure-temperature phase diagram of carbon according to Ref. [27] together with  pressure-temperature  conditions (rectangles) of synthesis of \emph{Pnnm} CN compound.} The products as determined by XRD measurements are noted inside the rectangles. The inset shows the formation enthalpies per formula unit, determined by theoretical calculations of this study, of the tetragonal \emph{P42/m}, orthorhombic \emph{Pnnm} and monoclinic \emph{Pc} CN phases relative to carbon and nitrogen as a function of pressure.}
\end{figure}

\begin{figure}[ht]
\includegraphics[width=120mm]{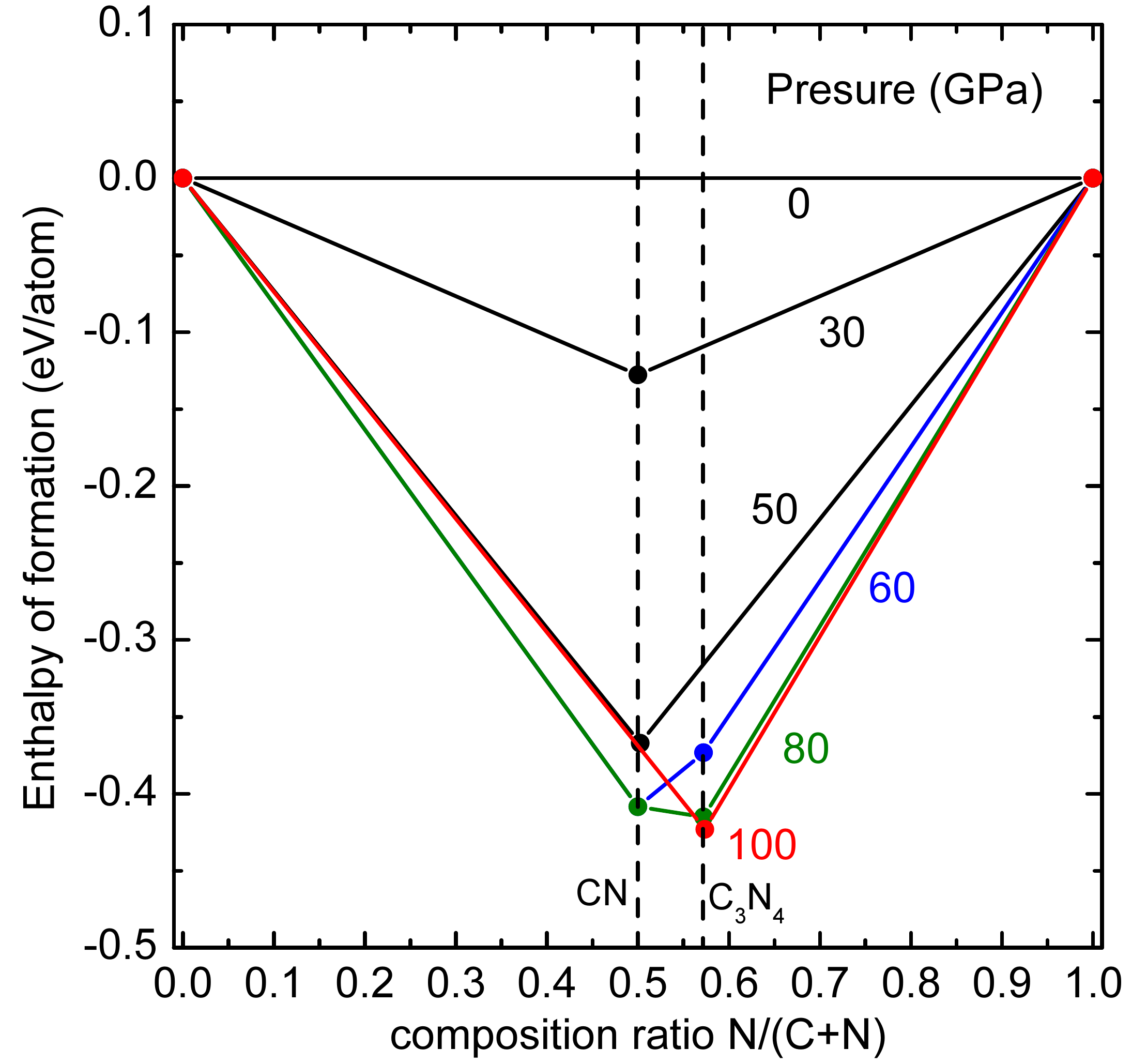}
\caption{\textbf{Predicted convex hull diagrams of C-N system at selected pressures.} Solid circles represent stable compounds.}
\end{figure}

\begin{table}[t]
  \caption{List of  experiments aiming to the CN synthesis together with thermodynamic conditions and products.}
  \centering
  \small
\begin{ruledtabular}
\begin{tabular}{l|l|l|l|l}

No.&	Pressure &	Temperature & 	Products at LH spot&	Comments \\
          & (GPa) & (K)&&\\
\hline
1	&$<$ 20	&Max $<$ 3000 	&C-Diamond	&Max T achieved \\
\hline
2	& 30 GPa	&Max $<$3500	&C-Diamond	&Max T achieved\\
\hline
3	&55 	& $>$ 7000	&\emph{Pnnm} CN	&Measured T before runaway\\
\hline
4	&65	&$>$7000	&\emph{Pnnm} CN	&Measured T before runaway\\
\hline
5	&$>$75 GPa	&4000$<$T$<$5000	&C-Diamond  	& Abrupt T raise $\approx$5000\\
&&&&Max T achieved\\

\end{tabular}
\end{ruledtabular}
\end{table}

\begin{table}[t]
  \caption{Miller-indices (\emph{h k l}), interplanar  distances (d$_{cal}$) and peak intensities calculated for the \emph{Pnnm} CN phase at 65 GPa, with unit cell parameters a=4.664\AA, b=3.664\AA {}  and c=2.455\AA, together with observed interplanar distances (d$_{obs}$).}
  \centering
  \small
\begin{ruledtabular}
\begin{tabular}{l|l|l|l|l}

\emph{h k l}&   d$_{cal}$ (\AA)&Intensity & d$_{obs}$ (\AA)&	d$_{cal}$-d$_{obs}$ \\
\hline
110	&2.886&	68&	2.887&	-0.001\\
200&	2.332	&75	&2.332&	0.000\\
101&	2.172	&82	&2.166&	0.006\\
011	&2.039&	56	&2.032&	0.007\\
210	&1.967&	30&	1.973&	-0.006\\
111&	1.869	&100	&1.869&	0.000\\
020	&1.832&	7&	1.83&	0.002\\
120	&1.705&	5&	1.691&	0.014\\
211	&1.536&	5&	1.529&	0.007\\

\end{tabular}
\end{ruledtabular}
\end{table}

 \begin{figure}[ht]
\includegraphics[width=150mm]{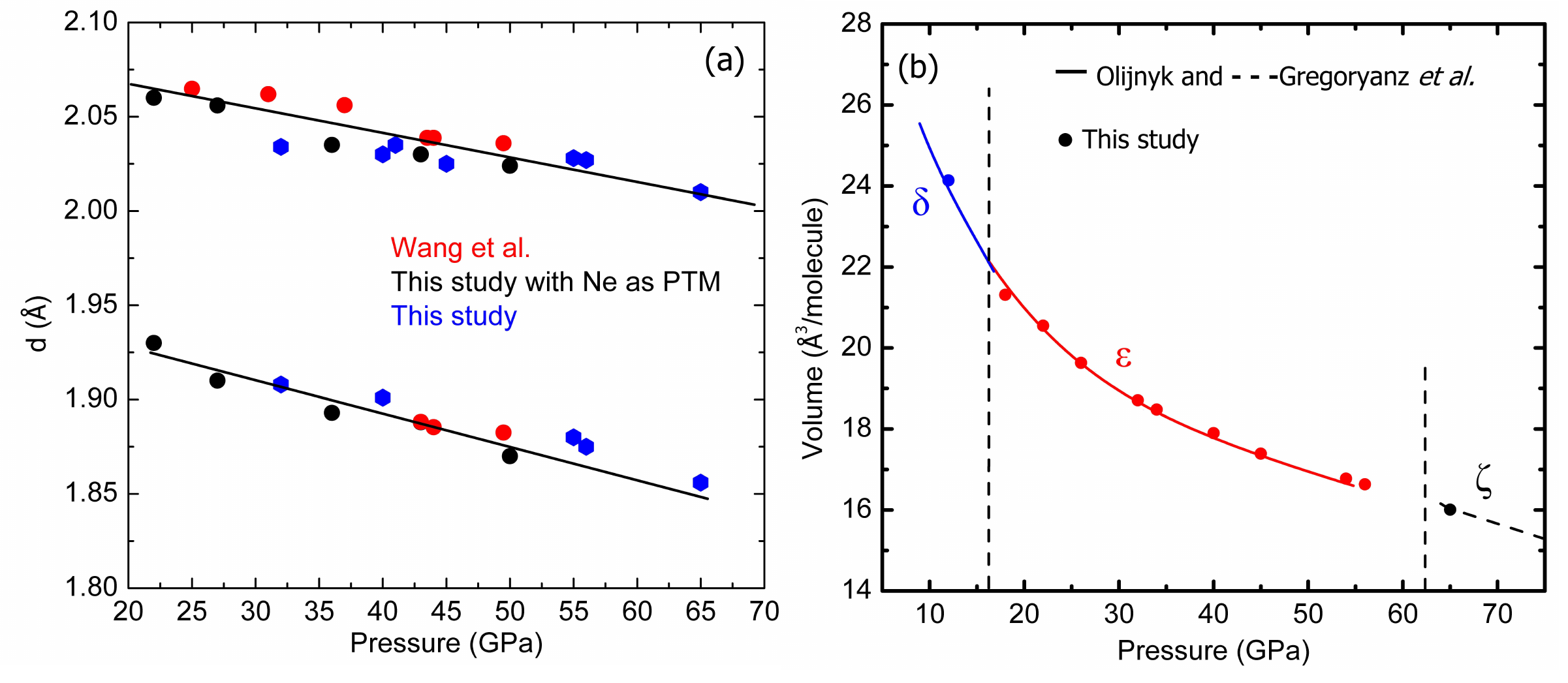}
\caption{\textbf{(a) d-spacing of the main peaks of the high pressure phase of carbon as a function of pressure:} Red symbols: Ref. [26] (starting from pure graphite with increasing pressure), Black symbols: this study (Ne as PTM) and Blue symbols: this study on pressure release. (b) \textbf{EOS of N$_2$,} solid lines from Refs. [23,24], solid circles this study.}
\end{figure}

 \begin{figure}[ht]
\includegraphics[width=150mm]{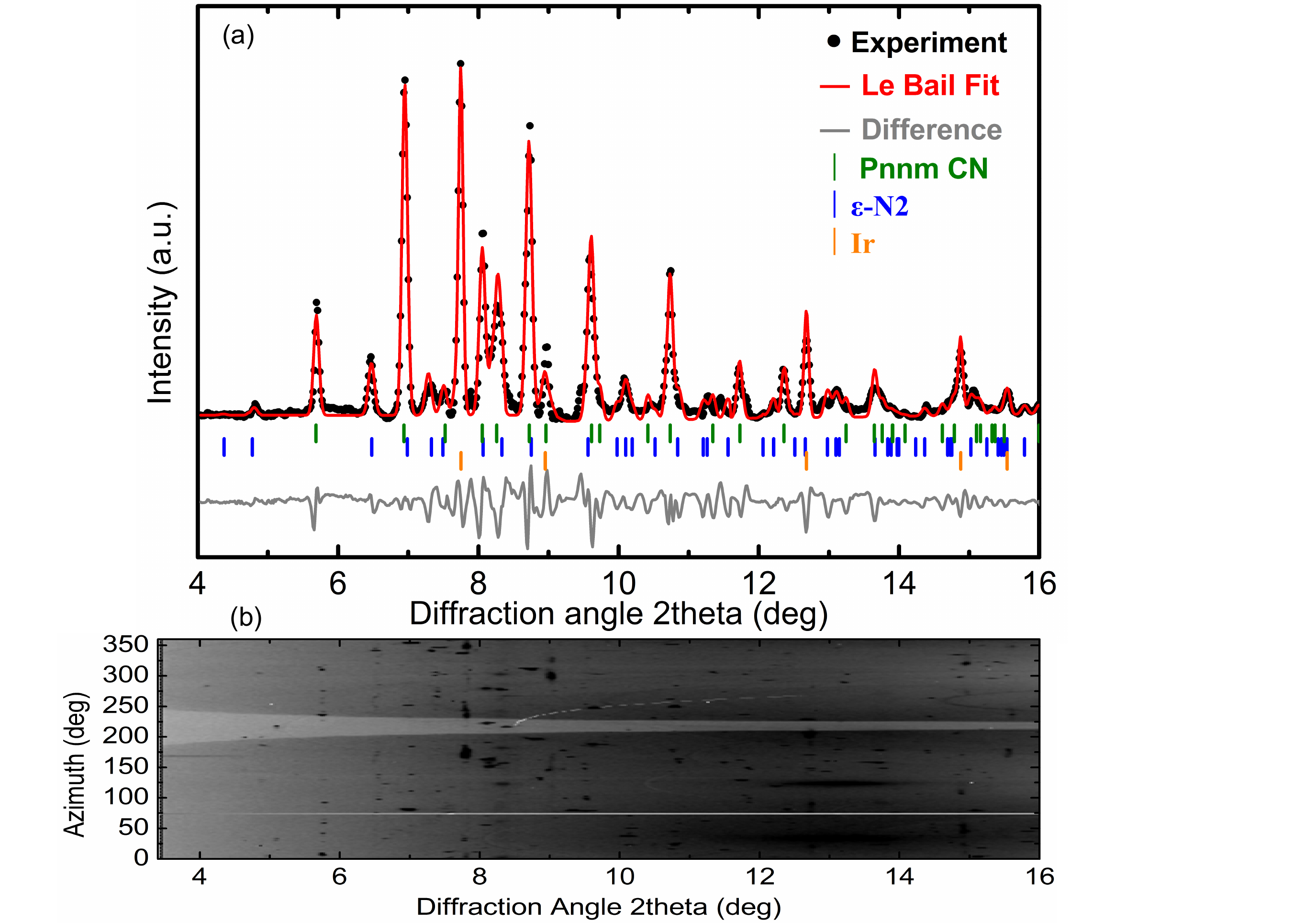}
\caption{ \textbf{(a) Le Bail refinement for CN at 55 GPa (first synthesis).} \emph{Pnnm} CN and $\epsilon$-N$_2$ peaks are marked with
green and blue vertical lines respectively. The X-ray wavelength is 0.2896 \AA. Although the X-ray beam was focused to the center of the Ir ring spacer, still the Bragg peaks of FCC iridium  appear due to the high Z in comparison to N and C and are marked with orange. We didn't observe any  sign of Ir-N compound formation moreover LH has been performed far away from Ir. (b) 2D X-ray diffraction image in rectangular coordinates (cake).}
\end{figure}

 \begin{figure}[ht]
\includegraphics[width=140mm]{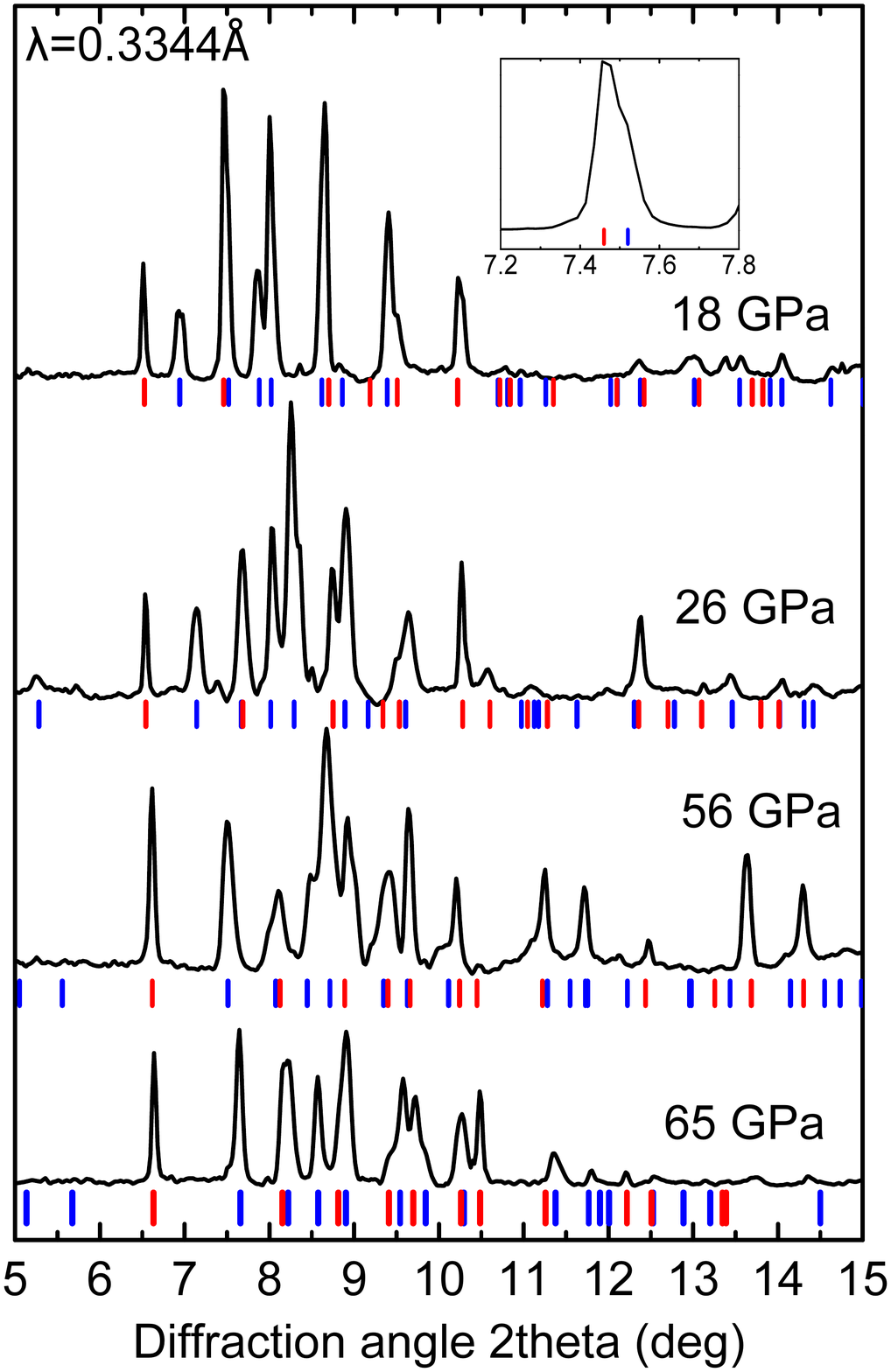}
\caption{\textbf{X-ray diffraction patterns of CN at various pressures on pressure release after synthesis at 65 GPa}. Expected positions of Bragg peaks of nitrogen and \emph{Pnnm} CN phase are indicated with vertical blue and red ticks respectively.}
\end{figure}

\begin{figure}[ht]
\includegraphics[width=140mm]{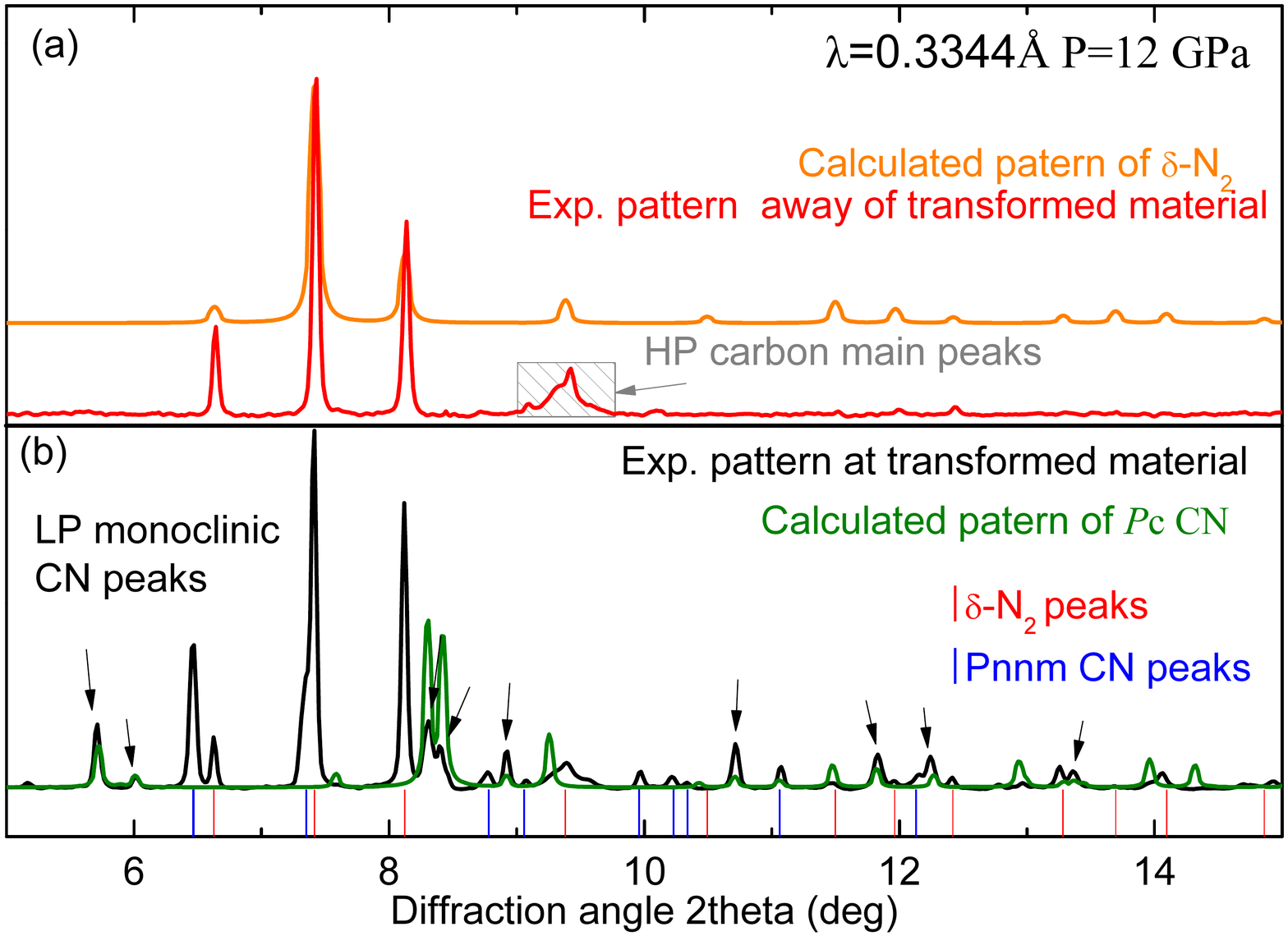}
\caption{\textbf{X-ray diffraction pattern at 12 GPa}: (a) Away from the synthesized material, the calculated pattern for $\delta$ phase of N$_2$ is plotted for clarity, (b) at the synthesized material, the calculated pattern for the LP CN monoclinic \emph{Pc} phase is plotted with green. The experimental and theoretical (in brackets) lattice parameters are: a= 4.607(4.5421)\AA, b= 3.172(3.2461) \AA, c= 3.615(3.5429) \AA, and beta= 110.85 (108.855)deg.  }
\end{figure}

\begin{figure}[ht]
\includegraphics[width=100mm]{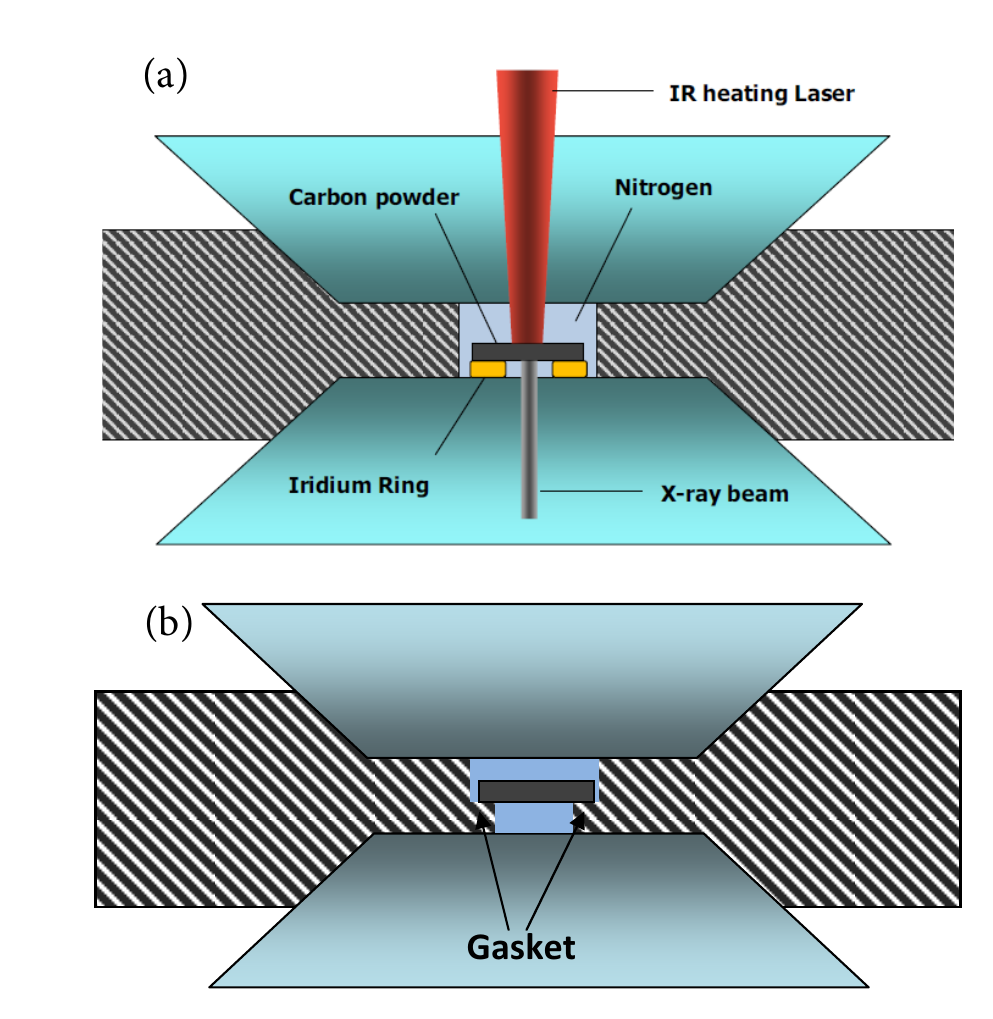}
\caption{\textbf{Schematic representation of the cross-section of the experimental arrangements:} a) Iridium ring spacer and b) recessed gasket. Double-sided LH has been performed but the schematic is simplified for clarity.}
\end{figure}

\end{document}